\documentclass[prl,twocolumn,superscriptaddress,preprintnumbers,floatfix]{revtex4-2}
\usepackage[T1]{fontenc}
\usepackage{epsfig}
\usepackage{graphicx}
\usepackage[english]{babel}
\usepackage{hyphenat}
\usepackage{amsmath}
\usepackage{amssymb}
\usepackage{mathtools}
\usepackage{mathrsfs}
\usepackage{slashed}
\usepackage{epstopdf}
\usepackage{xcolor}
\usepackage{braket}
\usepackage{booktabs}
\definecolor{lcolor}{rgb}{0.,0.0,0.}
\definecolor{citcolor}{rgb}{0,0.,0.5}
\usepackage[breaklinks,colorlinks,urlcolor=blue,citecolor=blue,linkcolor=blue]{hyperref}
\usepackage{multirow}
\usepackage{ltablex}
\usepackage{soul}

\usepackage[e]{esvect}

\newcommand{\secn}[1]{Section~1}
\newcommand{\appn}[1]{Appendix~1}

\long\def\comment#1{ }

\def\and{\quad\text{and}\quad}

\def\0{{\boldsymbol 0}}
\def\1{{\boldsymbol 1}}

\def\0{{\boldsymbol 0}}

\def\P{{\boldsymbol P}}

\def\bn{{\boldsymbol n}}

\renewcommand{\part}{{\rm part}}

\newcommand{\be}{\begin{equation}}
\newcommand{\ee}{\end{equation}}
\newcommand{\bes}{\begin{subequations}}
\newcommand{\ees}{\end{subequations}}
\newcommand{\bea}{\begin{eqnarray}}
\newcommand{\eea}{\end{eqnarray}}

\newcommand{\nn}{\nonumber \\}

\def\bea#1\eea{\begin{align}#1\end{align}}
\newcommand{\bef}{\begin{figure}[h!tb]\centering}
\newcommand{\eef}{\end{figure}}

\allowdisplaybreaks

\begin{document}

\title{Dissecting Jet Modification in the QGP with Multi-Point Energy Correlators}

\author{Jo\~{a}o Barata}
\affiliation{CERN, Theoretical Physics Department, CH-1211, Geneva 23, Switzerland}

\author{Ian Moult}
\affiliation{Department of Physics, Yale University, New Haven, CT 06511}

\author{Andrey V. Sadofyev}
\affiliation{
LIP, Av. Prof. Gama Pinto, 2, 1649-003 Lisbon, Portugal
}

\author{Jo\~{a}o M. Silva}
\affiliation{
LIP, Av. Prof. Gama Pinto, 2, 1649-003 Lisbon, Portugal
}
\affiliation{Departamento de Física, Instituto Superior Técnico (IST), Universidade de Lisboa, Av. Rovisco Pais 1, 1049-001 Lisbon, Portugal}

\affiliation{Departamento de Física Teórica y del Cosmos, Universidad de Granada, Campus de Fuentenueva,
E-18071 Granada, Spain}

\preprint{CERN-TH-2025-029}

\begin{abstract}
Energy correlators have recently attracted significant attention in the study of heavy ion collisions due to their potential to robustly connect experimental measurements with an underlying quantum field theoretic description. While theoretical studies have so far primarily focused on the simplest two-point correlator, mapping out the dynamics of the quark-gluon plasma (QGP) will require developing a theoretical understanding of multi-point energy correlators. In this paper we present a systematic theoretical study of multi-point energy correlators for jets fragmenting in a dense quark-gluon plasma, accounting for both the medium's perturbative modification to the jet, and its hydrodynamical back-reaction. 
We consider both the scaling behavior of projected correlators, as well as the shape dependent three-point correlator, highlighting how both provide insight into interactions with the QGP. 
We discuss the parametric dependence of modifications on the medium scales, opening new opportunities to experimentally separate jet modifications from the medium response. Our results open the door to a systematic exploration of multi-point energy correlators in heavy ion collisions.

\end{abstract}

\maketitle

\noindent\textbf{Introduction:} A primary goal of high-energy nuclear physics is the study of ordinary matter under extreme conditions, to search for emergent phenomena in the Standard Model. On Earth, high-energy heavy-ion collisions (HICs) are the prime experimental avenue to explore the finite temperature and density properties of QCD, allowing for the production of droplets of the quark-gluon plasma (QGP). Deciphering the underlying physics of the QGP from HICs requires the combination of different soft and hard probes to construct a tomographic picture of the emergent QCD matter, see e.g.~\cite{Busza:2018rrf} for a recent overview. Although a nearly comprehensive qualitative understanding of these events has been achieved, a significant knowledge gap remains in connecting various experimental findings to a first-principles description from the underlying theory of QCD.

A natural observable for achieving a direct connection between experimental measurements and QFT are matrix elements \cite{Basham:1978zq,Basham:1979gh,Basham:1977iq,Basham:1978bw} of the Average Null Energy (ANE) operator~\cite{Sveshnikov:1995vi,Tkachov:1995kk,Korchemsky:1999kt,Hofman:2008ar}:
\begin{align}\label{eq:def_Ec}
 \mathcal{E}(\bn) = \lim_{r\to \infty} r^{2
 } \int_0^\infty dt\, n^i\, T^{0i}(\bn)\, ,
\end{align}
which experimentally corresponds to an idealized calorimeter cell on a sphere of radius $r$ along the direction set by the normal vector $\bn$. Due to their operator definition in terms of the stress tensor of the underlying theory, one can hope to use symmetries \cite{Chen:2022jhb,Chang:2022ryc}, or the operator product expansion \cite{Andres:2024xvk} to make robust statements about their behavior in some nuclear environments.

Energy correlators were introduced as jet substructure observables in \cite{Basham:1978zq,Basham:1979gh,Basham:1977iq,Basham:1978bw,Korchemsky:1999kt,Dixon:2019uzg,Chen:2020vvp,Komiske:2022enw}, and first applied as an approach to study the QGP in \cite{Andres:2022ovj}. They have since been measured in proton-proton \cite{Komiske:2022enw,CMS:2024mlf,ALICE:2024dfl}, proton-nucleus \cite{talk_Anjali_fixed}
, and nucleus-nucleus \cite{CMS-PAS-HIN-23-004} collisions, and clear nuclear modification has been observed. This has prompted significant theoretical efforts to understand these observables, see e.g. \cite{Andres:2024ksi,Yang:2023dwc,Barata:2023bhh,Singh:2024vwb,Barata:2023zqg,Andres:2022ovj,Xing:2024yrb,Apolinario:2025vtx,Andres:2024xvk,Bossi:2024qho, Andres:2023xwr,Andres:2022ovj}.

In the HIC context, most of the discussion has been focused on the simplest two-point energy correlator.  Much in analogy with cosmology, while the two-point correlator is able to detect the presence of a scale, higher point correlators are able to probe the underlying dynamics. Higher point correlators will therefore be crucial in distinguishing different underlying properties of the QGP. Ultimately, we envision a program, much in analogy with cosmology, of mapping out the shapes of higher point correlators in different phases of QCD matter produced in HICs. 
In vacuum, the three- \cite{Chen:2019bpb} and four-point \cite{Chicherin:2024ifn} correlator have been computed, analyzed \cite{Chen:2022jhb} and measured \cite{Chen:2022swd}. An initial exploration of the three-point correlator in the QGP was presented in \cite{Bossi:2024qho}, highlighting its utility for mapping out the shape of the wake of jets. 
A related discussion of energy loss effects on higher-point correlators was presented in~\cite{Barata:2024bmx}.
To further 
expand our understanding of jet substructure in the QGP will require developing a theoretical understanding of multi-point correlators, and their modification by the QCD matter produced in HICs. 

While it may seem that a robust understanding of multi-point energy correlators in HICs is beyond, two recent developments provide optimism.  First, \cite{Andres:2024xvk} initiated a discussion of
the light-ray operator product expansion (OPE) to the study of energy correlators in  nuclear environments, showing on general symmetry grounds that medium modification could imprint  
itself with well defined scaling laws into the energy correlators.  Second \cite{Barata:2024ieg} showed how the energy correlator observables could be understood in hydrodynamical states, also leading to robust scaling laws. Combined, these two approaches provide the means to analyze both perturbative and hydrodynamic medium response of multi-point energy correlators.

In this \textit{Letter} we present the first theoretical calculation of higher-point energy correlators in the QGP. We consider both the full shape dependent three-point correlator (E3C), as well as the projected multi-point correlators (PENCs) \cite{Chen:2020vvp}. Accounting for both the medium induced perturbative modifications to the jet cascade, and the hydrodynamically driven back-reaction, we show how the QGP imprints itself onto the (P)ENCs' structure set by the symmetries of the $\mathcal{E}$ operators in the OPE limit. We argue that this mapping depends very weakly on the particular assumptions made in the present calculation and is underlined by universal features. Although such conclusions are natural from the perturbative side~\cite{Andres:2024xvk}, where we derive a closed form for the OPE coefficients, we show a non-trivial correspondence between the OPE limit and the medium response's structure, dictated by purely geometrical correlations~\cite{Barata:2024ieg}. More, the medium response and the perturbative contributions have distinct parametric dependencies on the medium's scales, allowing for a future first theory-driven approach to disentangle them.

\noindent\textbf{Projected Energy Correlators:} We begin by considering the simpler case of the projected energy correlators \cite{Chen:2020vvp}, which are obtained by integrating out the shape information, keeping only the information on the largest angle. The PENCs exhibit a scaling, which is sensitive to the twist-2 anomalous dimensions in the theory.

\begin{figure*}[t!]
    \centering
    \includegraphics[width=.28\textwidth]{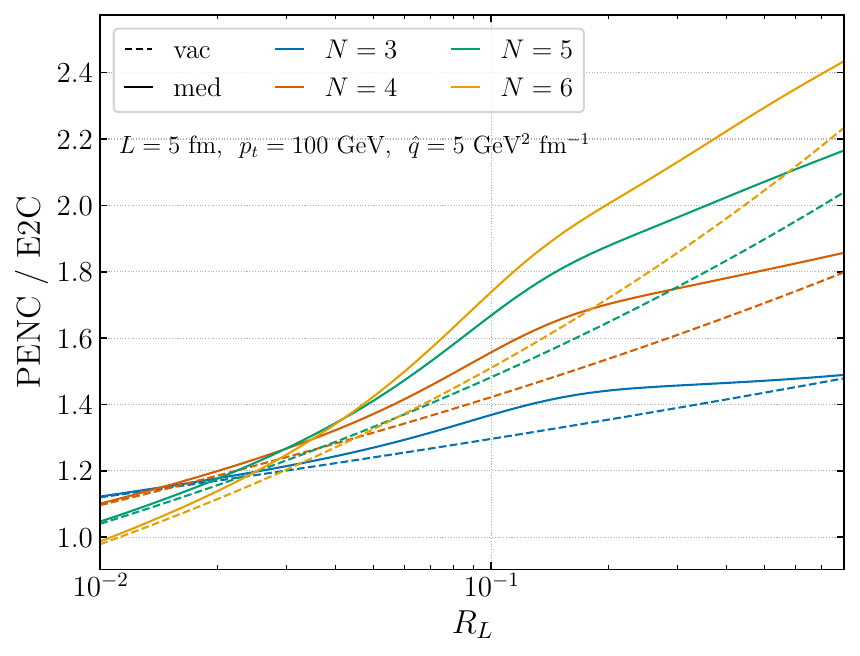} \hspace{.2 cm}
    \includegraphics[width=.28\textwidth]{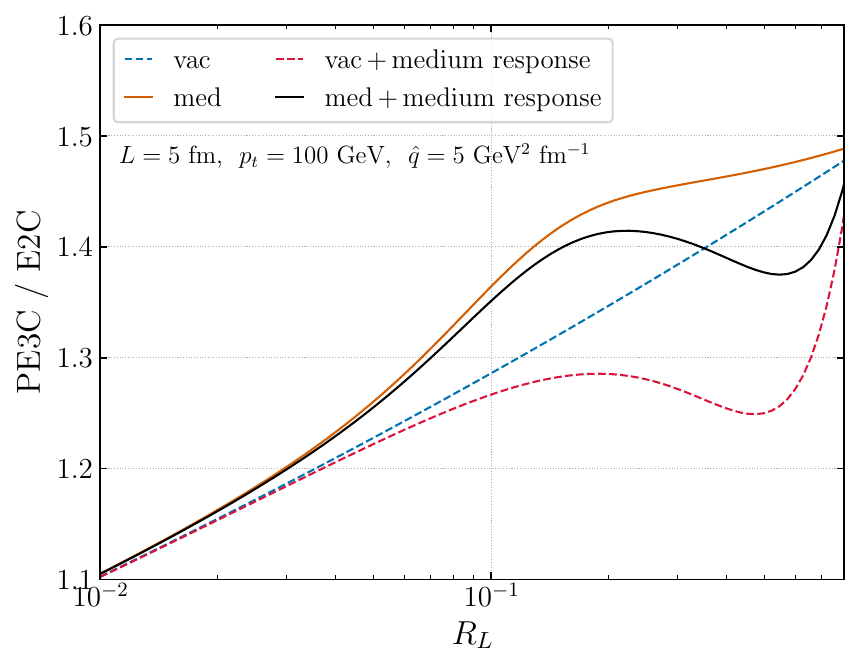}  \hspace{.2 cm}
    \includegraphics[width=.28\textwidth]{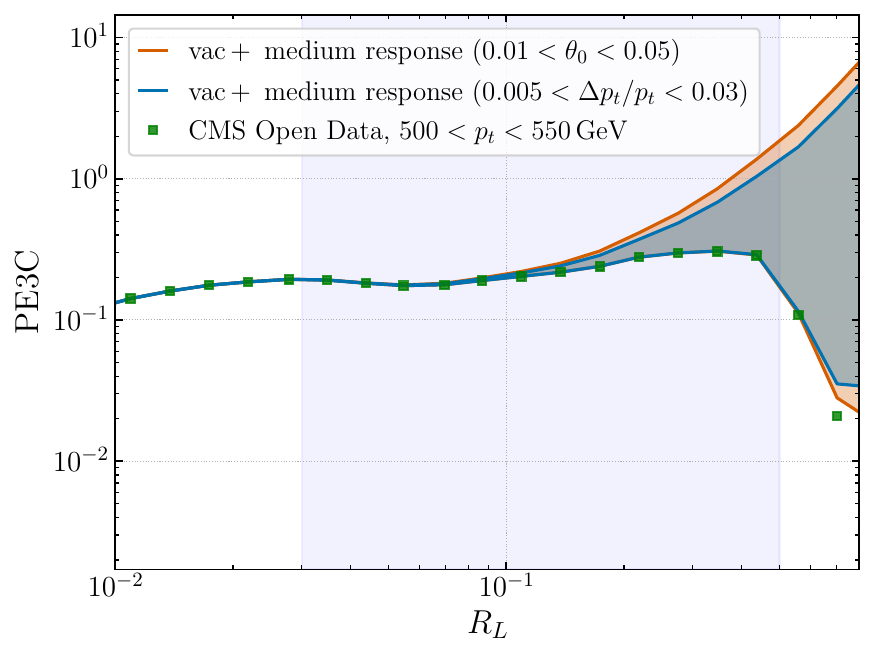}
     \caption{\textbf{Left:} PENC/E2C ratios for the vacuum (dashed) and in-medium (solid). \textbf{Centre:} PE3C/E2C including the medium response. \textbf{Right:} PE3C extracted from CMS Open Data presented in~\cite{Komiske:2022enw}, with a theory band for the medium response, obtained by convoluting with the E2/3C data. The $\theta_0$ band has $\Delta/p_t = 0.02$ while the $\Delta/p_t$ band has $\theta_0 = 0.025$.  The purple shaded area indicates the region where the perturbative calculation of the EEC is valid; the points for $R_L>0.5$ lie outside the cone of the jet where the PE3C is extracted.}  \label{fig:overall}
\end{figure*}

In the collinear limit and at leading accuracy, the PENC quark distribution, $d\Sigma_{\rm P}^{(N)}$, reads at fixed coupling~\cite{Chen:2020vvp}:
\begin{align}\label{eq:PENC_def}
 \frac{ d \Sigma^{(N)}_{\rm P}}{dR_L}  =   \frac{-\alpha_s  R_L^{ \frac{\alpha_s}{\pi}\gamma(N+1)} }{\pi R_L}  \int_0^1 dz \, P_{gq}
\sum_{k=1}^N \binom{N}{k} z^k
\, ,
\end{align}
with the anomalous dimension $\gamma(j)\equiv - \int_0^1 dz\, z^{j-1}\hat P_{gq}(z)$ of spin $j=N+1$, and $\hat P_{gq}$ the regularized splitting function. The PENC/E2C ratio gives direct access to the perturbative anomalous scalings in QCD, which we illustrate in Fig.~\ref{fig:overall} (\textbf{top}) in dashed lines for $N=3-6$.

In the conformal limit of QCD, the structure of the correlators is set by the symmetries of $\mathcal{E}$, allowing for an operator product expansion \cite{Hofman:2008ar}. 
In particular, this imposes that the PENC can be expanded about the collinear limit as,
\begin{align}\label{eq:hj1}
 \frac{ d \Sigma^{(N)}_{\rm P}}{dR_L}  = \sum_{k=1} a^{{\rm P}(N)}_{\tau,0} R_L^{\tau-3}\Big\vert_{\tau=2k}\,.
 \end{align}
Note that Eq.~\eqref{eq:PENC_def} satisfies this form at leading power.

 We first discuss how the QGP imprints onto Eqs.~\eqref{eq:hj1} and \eqref{eq:hj2}, focusing on
perturbative modifications to fragmentation and the medium back-reaction. We assume a static and homogeneous QGP, indicating when these assumptions become relevant. 

\begin{figure}[]
    \centering
    \includegraphics[width=.48\columnwidth]{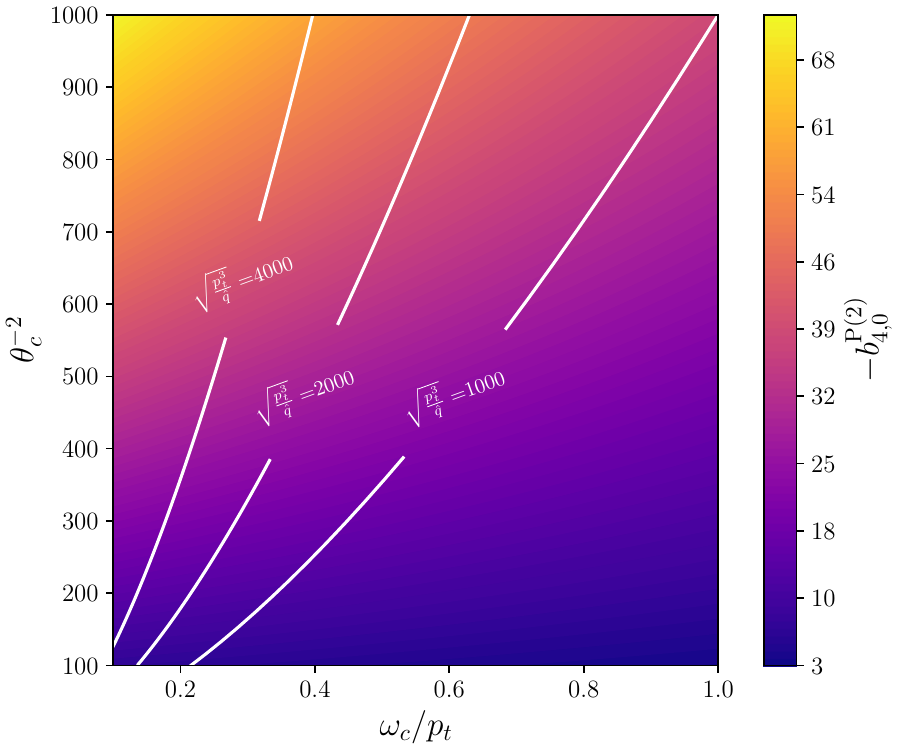}  \hspace{.0 cm}
     \includegraphics[width=.48\columnwidth]{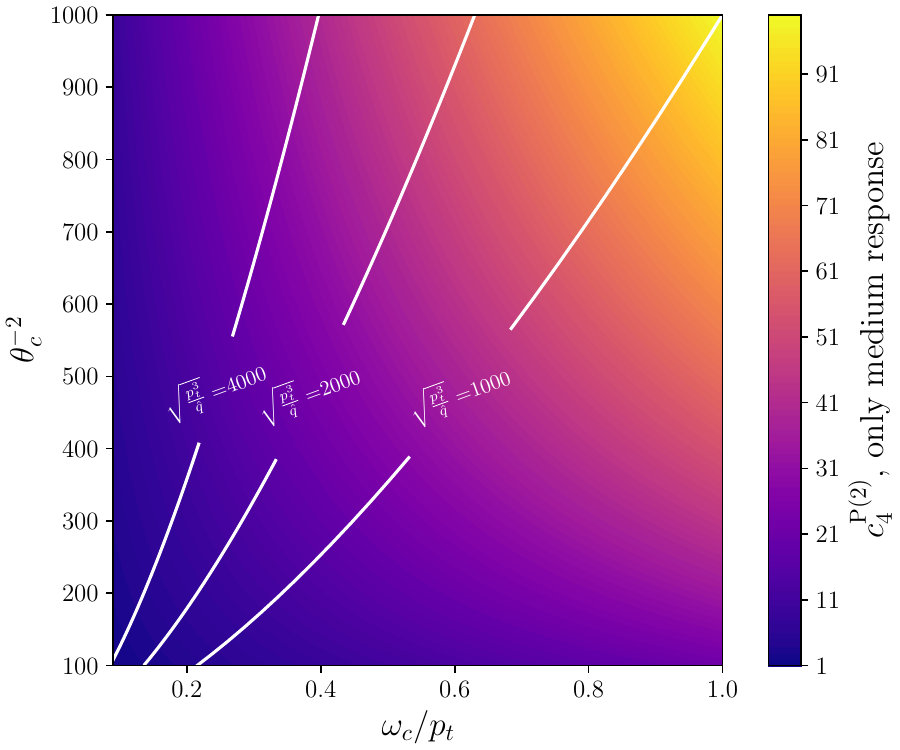}
     \caption{OPE limit coefficients for the medium perturbative corrections, $b_{4,0}^{{\rm P}(2)}$, and the response, $c_{4}^{{\rm P}(2)}$, neglecting hard correlations. White lines have $p_t^3/\hat q$ constant.}
    \label{fig:E3C_uu1}
\end{figure}

At leading order, the perturbative corrections can be absorbed into a  modification of the splitting function entering Eq.~\eqref{eq:PENC_def}. One can write $P_{gq}=P_{gq}^{\rm vac}+P_{gq}^{\rm med}$, where the medium induced splitting functions $P^{\rm med}$  have been widely discussed in the literature~\cite{Isaksen:2023nlr,Isaksen:2020npj,Sievert:2019cwq,Apolinario:2014csa,Blaizot:2012fh}, and are detailed in the supplemental material. Importantly, one can show that, for a static and homogeneous medium, $P^{\rm med}=P^{\rm med}(\omega_c,\theta_c)$, where $\omega_c\propto \hat q L^2$ is the critical gluon frequency below which the QCD LPM effect is prevalent and $\theta_c^{-2} \propto \hat qL^3$ the respective characteristic angle, with $L$ the medium longitudinal size, and $\hat q$ the jet quenching transport coefficient. Using the explicit form of $P^{\rm med}$, the PENC distribution in the OPE limit reads:
\begin{align}\label{eq:hj1_med}
 \frac{ d \Sigma^{(N)}_{\rm P, \,  med \, (no \, vac) }}{dR_L}  = \sum_{k=2} b^{{\rm P}(N)}_{\tau,0}(\omega_c/p_t,\theta_c) R_L^{\tau-3}\Big\vert_{\tau=2k} \, ,
 \end{align}
 where the coefficients $b^{{\rm P}(N)}_{\tau,0}(\omega_c/p_t,\theta_c)$ are given in the supplemental material, depending only on two parameters, with $p_t$ the jet energy. As an illustration, the leading OPE coefficient, $b_{4,0}$, is shown in Fig.~\ref{fig:E3C_uu1}, as a function of $(\theta,\omega_c)$; note that $a_{\tau,0}$ would be constant in a similar plot. The parametric dependence in Eq.~\eqref{eq:hj1_med} becomes more complex in an evolving medium, but can still be derived~\cite{Adhya:2020xcb,Baier:1998yf}. Critically, following typical jet quenching theoretical considerations, all odd twist coefficients vanish, since CP-odd interactions are neglected and the medium is structureless. Measurements yielding non-vanishing odd coefficients would then give access to non-perturbative features of the jet-medium interactions and the spatial structure of the QGP, not yet addressed by any other jet observable. Further discussion on these aspects is left to a future publication. In Fig.~\ref{fig:overall} (\textbf{left}) we show the perturbative medium contributions to the leading PENC ratios in solid lines. Compared to the vacuum, there is an enhancement increasing with $N$ at large $R_L$, reflecting the region where $b_{\tau,0}>a_{\tau,0}$. 

In turn, the imprint of the jet-induced medium response can be computed for any PENC. This follows from the fact that the response's ANE is purely classical, at least in all the present theoretical descriptions of the response, and thus the respective PENCs are dominated by geometrical correlations, as discussed in detail for the E2C in~\cite{Barata:2024ieg}. Restricting, for simplicity, the discussion to the PE3C/E2C ratio, 
 we split the ANE operator $\mathcal{E}(\bn) =\hat{\mathcal{E}}_{\rm h}(\bn)+\mathcal{E}_{\rm c}(\bn) $.
The first term represents the fluctuating part of the flow operator associated to the hard (h) jet fragmentation, just discussed above, while the second term accounts for the classical (c) flux, including the medium response. The corresponding contributions to the PE3C (ccc, chh, and hhh) can be expressed through the leading perturbative correlators of the h component combined with the c piece. Here, we do not attempt to constrain particular details of the classical energy flux in an event with a jet, which would go beyond the current theoretical control, but rather rely on a two-parameter model, $\mathcal{E}_c(\bn)=\frac{\Delta}{\pi\theta_0^2} e^{-\theta^2/\theta_0^2}$, where $\theta_0$ controls the angular size of the classical energy flux and $\Delta$ accounts for the amount of the energy which is propagating in the medium. It should be stressed that the exact functional form of $\mathcal{E}_c$ does not play an important role, and the classical flux is rather determined by the magnitude of the energy propagating in the matter, controlled by $\Delta$, and the respective characteristic angular region $\theta_0$. Omitting other medium-induced effects for the hard component and using
$\frac{d\Sigma^{(2,3)}_{\rm P, vac}}{dR_L}\simeq a^{P(2,3)}_{2,0} R_L^{\gamma(3,4)-1}$, where $3 \, a^{P(2)}_{2,0}=2 \, a^{P(3)}_{2,0}$ 
at the leading order, we find:
\begin{widetext}
\begin{align}\label{eq:ratio:E3C_E2C}
&\frac{\rm PE3C}{\rm E2C}\Bigg\vert_{\rm vac + response}{\hspace{-.5 cm}} = \Bigg(\frac{ a^{P(3)}_{2,0}}{R_L^{1-\gamma(4)}} 
+ \frac{ a^{P(3)}_{2,0} \Delta}{6\pi p_t}\,\mathcal{C}\,R_L^{1+\gamma(3)} 
+ \frac{\Delta^3}{\theta_0^2p_t^3}\,\left(\frac{2}{9}-\frac{\sqrt{3}}{6\pi}\right)R_L^3\Bigg)\Bigg(\frac{ a^{P(2)}_{2,0}}{R_L^{1-\gamma(3)}}+\frac{\Delta^2 R_L}{p_t^2\theta_0^2}\Bigg)^{-1} \, ,
\end{align}
\end{widetext}
where $\mathcal{C}$ corresponds to the chh terms and contains a collinear divergence numerically regulated at small angles, dominated by hadronization effects in two-point correlators, and, for simplicity, we assume that only the direction of the classical flux varies from event to event, while other fluctuations can be straightforwardly accounted for. An evaluation of Eq.~\eqref{eq:ratio:E3C_E2C}, comparing to the vacuum and the perturbative medium correction is given in Fig.~\ref{fig:overall} (\textbf{centre}), with $\theta_0=0.025$, $p_t=100$ GeV, $\theta_{\rm cut}=\Lambda_{\rm QCD}/p_t=0.002$, $\Delta = 0.01 \, p_t$,  $a^{P(3)}_{2,0} = 0.286$, fixing $\alpha_s = 0.3$. Including the response leads to a characteristic imprint in the ratio's shape, with a depletion at large angles, not observed when considering only the perturbative result. This is in qualitative agreement with recent Hybrid model results~\cite{Bossi:2024qho, Casalderrey-Solana:2014bpa,Casalderrey-Solana:2015vaa,Casalderrey-Solana:2016jvj}. To 
better gauge the phenomenological impact of these contributions, in Fig.~\ref{fig:overall} (\textbf{right}) we combine the medium response with the CMS data analysis in~\cite{Komiske:2022enw}; resulting in a large angle enhancement of the PE3C, comparable to the perturbative counterparts. {We reiterate} that the shape of PE2/3C depends weakly on the functional form of the classical flux, although being sensitive to the parameters taken. Similarly to the perturbative contributions, we provide the leading response coefficient $c_{4}$ in Fig.~\ref{fig:E3C_uu1} as a function of $(\theta,\omega_c)$, displaying the different scaling of the response with respect to the perturbative terms.

In general, the E2C associated to the medium response, in the collinear limit, takes the form 
\begin{align}\label{eq:hj1_wakes}
 \frac{ d \Sigma^{(2)}_{\rm P, \,  response}}{dR_L}  = \sum_{k=2} c^{{\rm P}(2)}_{\tau}\Big[ {\mathcal{E}}_c\Big] R_L^{\tau-3}\Big\vert_{\tau=2k} \, ,
\end{align}
where the coefficients are functionals of the classical energy flux, given by angular averages of its powers and derivatives. Remarkably, the expansion in Eq.~\eqref{eq:hj1_wakes}, independent of the OPE for the ANE, has the same structure as Eqs.~\eqref{eq:hj1} and \eqref{eq:hj1_med}. More, the leading coefficients in the expansion for the response's PE3C also follow the OPE limit structure, see numerator of Eq.~\eqref{eq:ratio:E3C_E2C}; coefficients again only depend on ${\mathcal{E}}_c$. These relations, i.e. Eqs.~\eqref{eq:ratio:E3C_E2C} and \eqref{eq:hj1_wakes}, hold for any $N$ for the leading $\tau$ terms, and we conjecture they hold for all $\tau$, assuming the classical energy flux to be cylindrically symmetric, ignoring azimuthal fluctuations. However, we notice that such asymmetries may appear due to the medium evolution and anisotropy both in the in-medium splitting function and medium response, which can be especially important in small HIC systems~\cite{Barata:2023zqg,Sadofyev:2021ohn, Barata:2023qds, Kuzmin:2023hko,Barata:2024bqp,Kuzmin:2024smy}. These lead to finite even contributions to Eq.~\eqref{eq:hj1_wakes}, see the supplemental material, suggesting prospective avenues to disentangle the in-medium physics from vacuum QCD, and providing a tool to access the azimuthal jet substructure. 

Finally, all these observations explicitly show that the perturbative terms and the medium response have direct competing contributions to the PENCs, requiring a systematic analysis to be disentangled. Eqs.~\eqref{eq:hj1_med} and~\eqref{eq:hj1_wakes} provide the full parametric dependence of the coefficients in the OPE limit, indicating the path for a theoretically driven decomposition of these effects in experiment.

\noindent\textbf{Shape Dependent Three-Point Correlator:} We now consider the full shape dependent three-point correlator. In the collinear limit, it can be obtained by integrating the $1\to 3$ splitting functions~\cite{Chen:2019bpb,Chen:2022jhb} 
\begin{align}\label{eq:J3}
 J_3 = \int d\Phi_3 \, \frac{g^4 \, P_3 \, \,  z_1 z_2 z_3}{2s_{123}^2}\Big\vert_{\sum_{i} z_i=1}^{ s_{123}=\sum_{i\neq j} s_{ij}} \, , 
\end{align} 
where $(4\pi)^5d\Phi_3 =ds_{12}ds_{13} ds_{23}\prod_{i}  dz_{i} \, 4 (-D)^{-1/2} $ denotes the three-body phase space~\cite{Gehrmann-DeRidder:1997fom}, $z_i \equiv E_i/p_t$ is the energy fraction carried by the final state particle $i=1,2,3$, and flavor dependence is  implicit. The momentum variables read $s_{ij} =  z_i z_j p_t^2 \theta_{ij}^2 $, while $ D =  (z_3 s_{12}-z_1 s_{23}- z_2 s_{13})^2-4 z_1 z_2 s_{13}s_{23} <0$. We shall restrict the discussion to the quark channel, where the vacuum splitting kernels are $P_{2, \, {\rm vac}}=P_{gq}$, $P_{3, \,{\rm vac}}  = C_F^2 P_{3, \,{\rm vac}}^{\rm Abel.} + C_F C_A P_{3, \,{\rm vac}}^{\rm non-Abel.}$~\cite{Catani:1998nv}, with the Abelian and non-Abelian terms given in the supplemental material.

Since our theoretical description of $P_{3, \, {\rm med}}$ in HICs is still limited, \footnote{See however~\cite{Fickinger:2013xwa} for calculations of $P_{3, {\rm med}}$ in dilute matter} in this work we also consider a factorized \textit{cascade} approximation~\cite{Fickinger:2013xwa}, where the $1\to 3$ process is given by a succession of $1\to 2$ branchings:
\begin{align}\label{eq:P3_cascade}
\frac{P_{3, \,{\rm vac}}^{\rm \, cascade}}{s_{123}} &=  \frac{P^{2,1+3}_{gq}P^{1,3}_{gq}  }{s_{13}} + (1\leftrightarrow 2)
+\frac{P_{gq}^{1+2,3}P_{gg}^{1,2} }{s_{12}}  \, ,
\end{align}
where we used the spin averaged $1\to 2$ splitting functions, and employed the notation, e.g. $P_{gq}^{1,2}(x)\equiv  C_F (1+(1-x)^2)/x$, and $ P_{gg}^{1,2} =  2C_A \left(\frac{x}{1-x} + \frac{1-x}{x} + x(1-x)\right)$, with $x = \frac{E_1}{E_1+E_2} $. The mismatch between the exact result and the cascade approximation, at the E3C level, is mild. We verified the cascade form does not interfere with the medium effects discussed below, which are more prominent, and emerge in different sectors in the $(\xi,\phi)$ plane, where $\xi = \frac{\theta_{13}}{\theta_{23}}$, $ \sin^2 \phi = 1- \frac{(R_L-\theta_{23})^2}{\theta_{13}^2}$, with $0<2 \phi<\pi$ and $0<\xi<1$ in the region $R_L=\theta_{12}>\theta_{23}>\theta_{13}$. The deviations between the cascade and exact solutions are mainly constrained to regions where the three detectors are either in flattened or squeezed configurations, with a $\mathcal{O}(30\% - 50\%)$ magnitude; further details on this comparison can be found in the supplemental material. It will be interesting to extend this calculation using the full medium modified $1\to 3$ splitting functions \cite{Fickinger:2013xwa}.

\begin{figure}[h!]
    \centering
    \includegraphics[width=1\columnwidth]{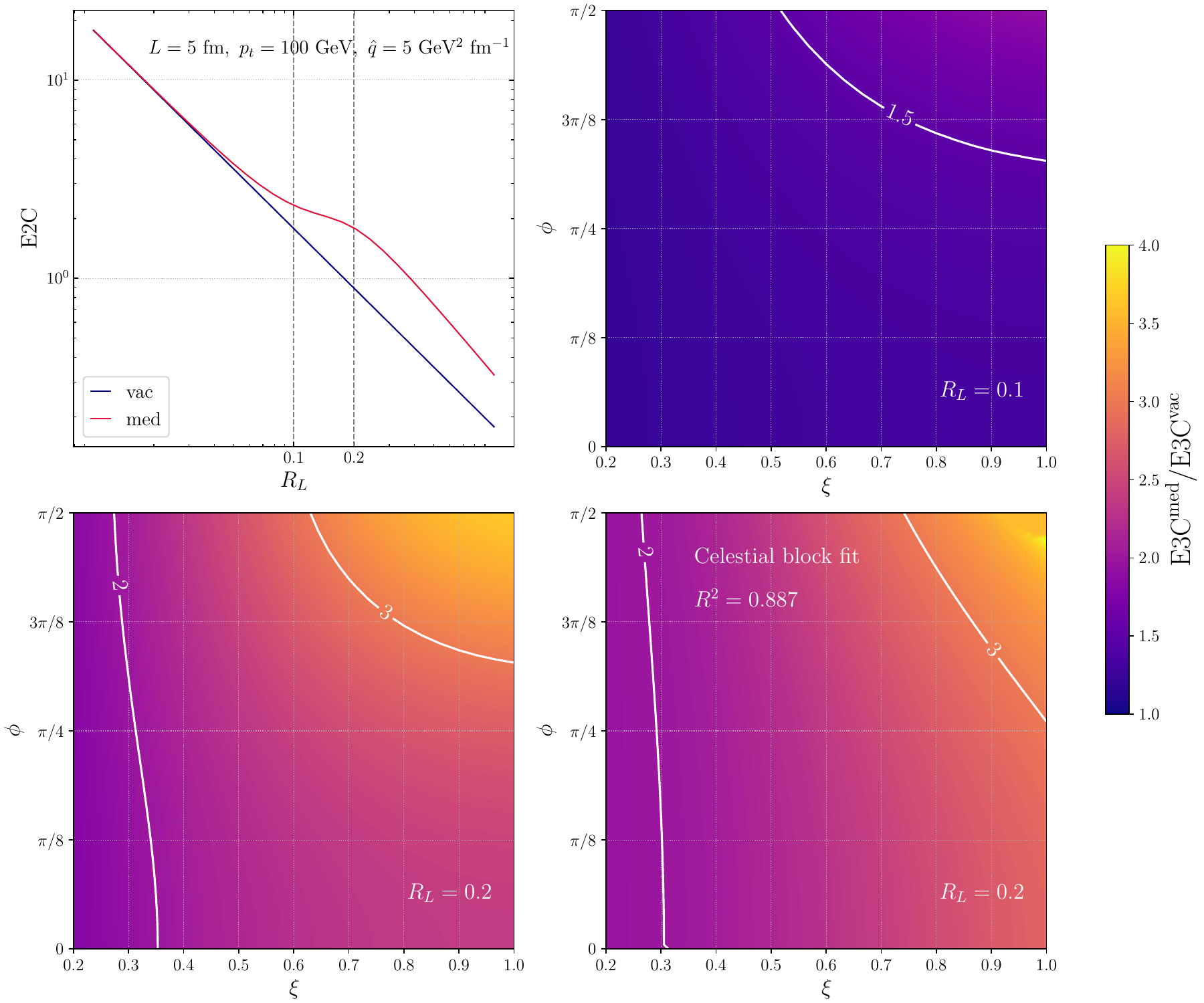}

    \caption{ 
    E2C distribution (\textbf{top left}) indicating the $R_L$ values used for the E3C/E2C ratios in-medium (\textbf{top right}, \textbf{bottom left}). Fit to the E3C/E2C ratio with $R_L=0.2$  (\textbf{bottom right}), following the celestial block structure.}

    \label{fig:2}
\end{figure}

The leading perturbative medium induced correction to $J_3$ is obtained straightforwardly, generalizing the cascade form in Eq.~\eqref{eq:P3_cascade} to the medium. The ratio of the in-medium and the vacuum E3C distributions is shown in Fig.~\ref{fig:2}, for two values of $R_L$ chosen by examining the respective E2C, as illustrated. The medium effects lead to a characteristic enhancement in the region where $R_L\approx \theta_{13}\approx \theta_{23}$. At larger $R_L$ the enhancement extends to lower values of $\phi$ at large $\xi$. The cascade approximation only mildly affects this sector.

The three-point energy correlator can be expanded in celestial blocks~\cite{Chang:2022ryc,Chen:2019bpb,Chen:2022jhb}:
\begin{align}\label{eq:hj2}
\, \frac{ R_L \theta_{23}\theta_{13}  \, d\Sigma^{(3)}}{\sqrt{-D}\, dR_L d\theta_{13} d\theta_{23}} = \sum_{k=1} a^{(3)}_{\tau,0} \, G^q_{\tau,0}(z,\bar z)\Big\vert_{\tau=2k} \, ,
\end{align}
where the complex variables $z$, $\bar z$, are defined as $R_L^2 z\bar z= \theta_{13}^2$, $R_L^2(1-z)(1-\bar z) = \theta_{23}^2$, while the single-valued $G^q_{\tau,0}$ can be written in terms of hypergeometric functions, whose explicit form can be found in \cite{Chen:2022jhb}. We fitted the data for $R_L=0.2$ to the functional form 
$\frac{\sum_{i=1}^3 n_i G_{2i,0}(\phi,\xi)}{\sum_{i=1}^3 d_i G_{2i,0}(\phi,\xi)}$, following the structure in Eq.~\eqref{eq:hj2}. This is shown in the bottom right plot of Fig.~\ref{fig:2}. We obtained a good fit to data, with $R^2=0.887$, while we find that the numerator's coefficient, for $i>1$, are approximately an order of magnitude larger than the denominator's ones. This points to the dominance of higher-twist terms in-medium, as expected. It will be interesting to explore this in more detail in different models.

\begin{figure}[t!]
    \centering
    \includegraphics[width=.28\textwidth]{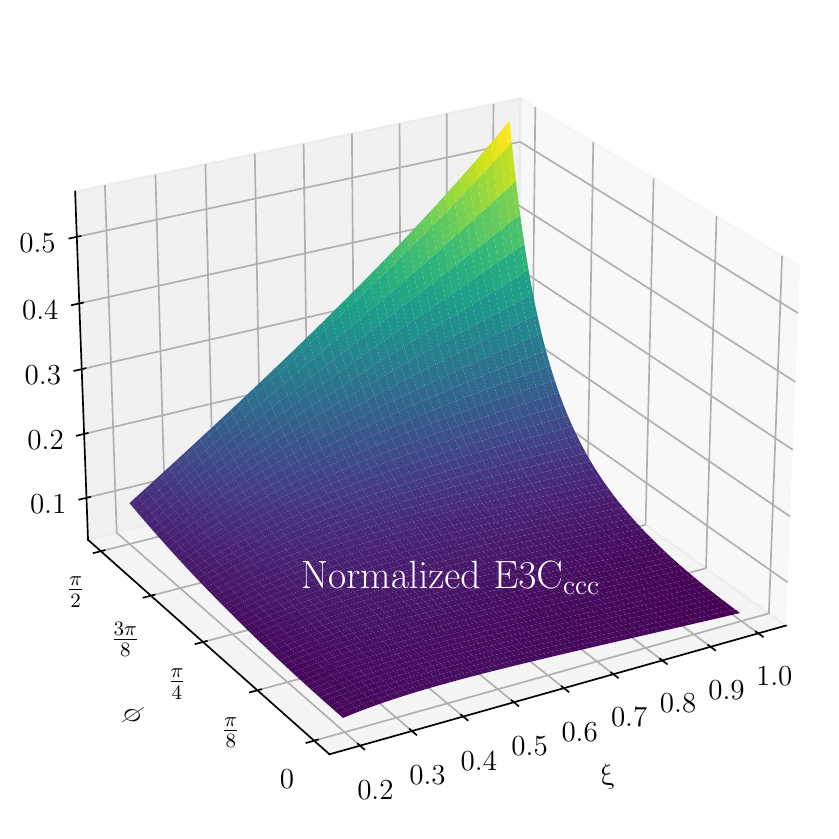} 
     \caption{Normalized E3C for the ccc term from hydrodynamic response.}
   \label{fig:E3C_final_plots}
\end{figure}

The E3C's extraction including the full response is far from trivial; nonetheless, insight into the distribution's shape can be gained from the ccc term:
\begin{align}
\nonumber
\frac{d\Sigma^{(3)}_{\rm ccc}}{dR_L d\xi d\phi} = \frac{ 4 \pi^2 p_t^{-2}R_L^3\xi\,\int d\theta_1 \, \mathcal{E}^3(\cos\theta_1)\sin\theta_1}{(1+\xi\cos\phi)^3\sqrt{4+4\xi\cos\phi-\xi^2\sin^2\phi}} \, ,
\end{align}
shown in Fig.~\ref{fig:E3C_final_plots}. The distributions' maximizer is nearly identical to the one for the perturbative terms in Fig.~\ref{fig:2}, further evidencing the overlap of these two effects, as for the PENCs. More, this result closely matches the numerical result in~\cite{Bossi:2024qho}; here it is solely determined by geometrical correlations. Finally, writing the ccc term in $(z,\bar z)$ coordinates, $
\frac{d\Sigma^{(3)}_{\rm ccc}}{dR_L d{\rm Re} z\,  d{\rm Im}z} = \frac{2\pi^2 R_L^3}{p_t^2}\left[\int d\theta_1 \, \mathcal{E}^3(\cos\theta_1)\sin\theta_1\right]$,
and decomposing it in the celestial blocks, we find in the (collinear) limit $z\to 0$ at fixed $\bar z$, the only contribution comes from the leading spin and $\tau=4$ block, demonstrating the overlap with higher terms in the twist expansion.

\noindent\textbf{Conclusion: } In this \textit{Letter}, we have presented the first theory study of the collinear limit of higher-point energy correlators in HICs, including both the perturbative modifications to the jet and the medium response. For the PENCs we have demonstrated the universal structure 
\begin{align}
\nonumber
   \frac{   d \Sigma^{(N)}_{\rm P}}{dR_L}  &= \sum_{\tau} \left\{a^{{\rm P}(N)}_{\tau,0} R_L^{- \delta_{\tau,2} \gamma(N+1)}+b^{{\rm P}(N)}_{\tau,0}+c^{{\rm P}(N)}_{\tau} \right\}R_L^{\tau-3}  \, ,
\end{align}
for even $\tau$, with $b^{{\rm P}(N)}_{2,0}=c^{{\rm P}(N)}_2=0$, assuming an azimuthal trivial background and jet substructure; odd $\tau$ terms can be associated to non-perturbative effects or non-trivial QGP profiles. Further, we have separated the parametric dependence of the different coefficients, permitting a first theory driven identification of the PENC structure. In Fig.~\ref{fig:E3C_uu1} we schematically show how to distinguish these elements, which exhibit distinct evolutions along (white) lines of constant $\hat q/p_t^3$. Experimental proxies for such curves would allow to map the PENCs in the QGP, see also e.g.~\cite{Mehtar-Tani:2024jtd}, a necessary and critical step for a clear interpretation of the experimental data.

At the level of the E3C, we have analytically shown that the perturbative and the medium response contributions overlap, as for the E2C case, and can not be directly distinguished apart, requiring a treatment analogous to the PENCs. More, we have explained the nature and shape of the medium response imprint on the E3C, previously only observed in numerical simulations~\cite{Bossi:2024qho}, clarifying their physical origin. We find strong evidences that the E3C follows the pattern set by the celestial block decomposition, with the QGP perturbatively imprinting itself on the higher-twist terms, with the response following the same structure. This further exposes the universal character of energy correlator observables in the QGP, complementing the recent discussions on the properties of these objects e.g.~\cite{Liu:2024lxy,Chen:2024nyc,Barata:2024wsu,Kang:2024dja,Alipour-fard:2024szj}.

\textit{Acknowledgment:} We are grateful to M. V. Kuzmin and J. G. Milhano for multiple fruitful discussions on related topics. We thank J. Brewer, W. Ke, K. Lee, U. A. Wiedemann for useful comments. This work is supported by European Research Council project ERC-2018-ADG-835105 YoctoLHC. The work of AVS is supported by Fundação para a Ciência e a Tecnologia (FCT) under contract 2022.06565.CEECIND. The work of JMS is supported under FCT project CERN/FIS-PAR/0032/2021 (http://doi.org/10.54499/CERN/FIS-PAR/0032/2021). The work of JMS has also been partially supported by MCIN/AEI (10.13039/501100011033) and ERDF (grant PID2022-139466NB-C21) and by Consejería de Universidad, Investigación e Innovación, Gobierno de España and Unión Europea – NextGenerationEU under grant AST22\_6.5. I.M. is supported by start up funds from Yale University, and by the Sloan Foundation.

\bibliographystyle{bibstyle.bst}

\bibliography{EEC_ref.bib}

\clearpage
\onecolumngrid

\section*{Supplemental material}

\subsection{Next to leading order vacuum splitting functions}
The $q\to q gg$ splitting function considered in the main text explicitly reads~\cite{Catani:1998nv}: 
\begin{align}
 & P_{3, \,{\rm vac}}  = C_F^2 P_{3, \,{\rm vac}}^{\rm Abel.} + C_F C_A P_{3, \,{\rm vac}}^{\rm non-Abel.}\, ,\nn 
 & P_{3, \,{\rm vac}}^{\rm Abel.} = \frac{s_{123}^2}{2s_{13}s_{23}} \frac{z_3(1+z_3)^2}{z_1z_2} + \frac{s_{123}}{s_{13}} \frac{z_3(1-z_1)+(1-z_3)^2}{z_1z_2} - \frac{s_{23}}{s_{13}} + (1\leftrightarrow 2) \nn 
 &P_{3, \,{\rm vac}}^{\rm non-Abel.} = \frac{(2(z_1s_{23}-z_2s_{13}) + (z_1-z_2)s_{12})^2}{4(z_1+z_2)^2s_{12}} +\frac{1}{4} + \frac{s^2_{123}}{2s_{12}s_{13}} \left(\frac{1+z_3^2}{z_2} + \frac{1+(1-z_2)^2}{1-z_3}\right)\, , \nn 
 &- \frac{s_{123}^2 z_3 (1+z_3^2)}{4s_{13}s_{23} z_1z_2} + \frac{s_{123}}{2s_{12}} \left( \frac{z_1(2-2z_1+z_1^2)-z_2(6-6z_2+z_2^2)}{z_2(1-z_3)}\right) \nn 
 &+ \frac{s_{123}}{2s_{13}} \left(\frac{(1-z_2)^3+z_3^2-z_2}{z_2(1-z_3)} -\frac{z_3(1-z_1)+(1-z_2)^3}{z_1z_2} \right) +  (1\leftrightarrow 2)\, ,
\end{align}
where in the collinear limit we recall that the invariants take the form $s_{ij} =  z_i z_j p_t^2 \theta_{ij}^2 $, where $z_i = E_i/p_t$, and $s_{123}=s_{12}+s_{13}+s_{23}$.

\begin{figure}[h!]
    \centering
    \includegraphics[width=.55\columnwidth]{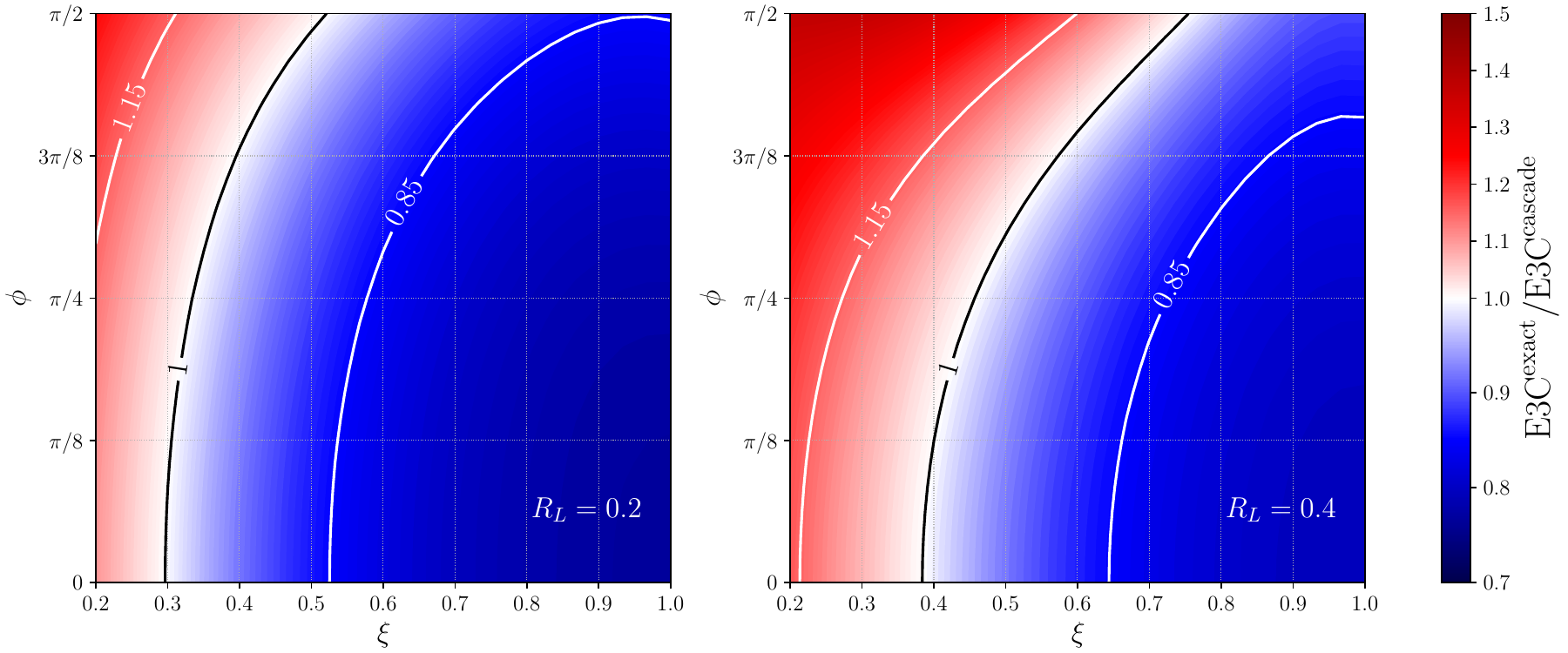}
    \caption{Leading order vacuum E3C ratio with respect to the cascade approximation.}
    \label{fig:cascase}
\end{figure}

Finally, we show in Fig.~\ref{fig:cascase} the ratio of the exact leading order E3C to the cascade approximation. The difference between these approximations are smaller than the medium effects we consider, ensuring the validity of the conclusions in the main text.

\subsection{Leading and next to leading order medium splitting functions}
Denoting the flavor dependence of each process as $i \rightarrow j k $, the medium induced perturbative kernels (with vacuum subtracted) are directly obtained from the respective particle distributions, see e.g.~\cite{Barata:2024bqp,Isaksen:2023nlr, Barata:2025future}:
\begin{align}\label{eq:dN_full_iso}
   &  \frac{2\pi}{\alpha_s}\P^2\frac{dN_{i\rightarrow jk}}{dzd\P^2} = C_{ji}P_{ji}(z)\Bigg(\frac{4\P^2}{\tilde p_t}\,{\rm Re} \int_{X_v^+} (-i)\frac{h_1^2}{h_3}e^{i\frac{\P^2}{2h_3\tilde p_t}}\left(1 + \frac{h_2}{h_3} + \frac{i\P^2}{2\tilde p_t}\frac{h_2}{h_3^2}\right) -2 {\rm Re} \left(1-e^{-i\frac{\P^2}{2\Omega\tilde p_t}\tan{(\Omega L^+)}}\right)\Bigg)\, ,
\end{align}
where $P_{ji}(z)$ is the massless $i \rightarrow jk $ vacuum splitting function stripped of its color factor, $z$ the light-cone energy fraction of parton $j$ with respect to parton $i$, $p_t$ the energy of the initial parton $i$ and $\tilde p_t = z(1-z)p_t \sqrt{2}$; the integration range is set by $X_v^+ = \{0 < x_v^+ < L^+, x_v^+ < \bar x_v^+ < L^+=L\sqrt{2}\}$. Note that the calculation is drastically simplified by assuming the independent broadening of the two final partons in the process, i.e., taking only the factorizable piece, see e.g.~\cite{Blaizot:2012fh,Apolinario:2014csa,Isaksen:2023nlr}. It was shown in~\cite{Isaksen:2023nlr} for $\gamma\rightarrow q\bar q$ that, when compared  with the finite-$N_c$ result, the large-$N_c$ result with only the factorizable piece quantitatively describes well the kinematic region where $|\P| =z(1-z) p_t \theta \lesssim \sqrt{\hat q L}$, while reasonably describing the main qualitative features of medium modifications outside of this region. The remaining relevant variables are defined as
\begin{align}\label{eq:ci_definition}
    & h_{1} = \frac{\Omega}{2i\sin{\Omega\Delta t}}\,, \qquad h_{2} = \frac{\Omega}{\tan{\Omega\Delta t}}\,, \qquad  h_{3} = \Delta_L^+\left(-i\frac{\hat q \, Q_{ji}(z)}{4\tilde p_t}\right)-h_{2}\,, \nn
    & \Delta_L^+ = L^+ - \bar x_v^+\, ,\qquad \Delta t = \bar x_v^+ - x_v^+\,, \qquad \Omega = \frac{1-i}{\sqrt{2}}\sqrt{\frac{\hat q F_{ji}(z)}{4\tilde p_t}}\,.
\end{align}
The process-dependent functions for the relevant splittings in this work read
\begin{itemize}
    \item $q \rightarrow g q$
    \begin{align}
    & C_{gq} = C_F, \,\, Q_{gq}(z) = 2(1-z)^2 + z^2,\,\,F_{gq}(z) = z P_{gq}(z)\,,
    \end{align}
    \item $g \rightarrow gg$
    \begin{align}
    & C_{gg} = 2C_A, \,\, Q_{gg}(z) = 2((1-z)^2 + z^2),\,\, F_{gg}(z) = 1 + z^2 + (1-z)^2\, .
    \end{align}
\end{itemize}
Having described the form of the $1\to 2$ splitting kernels, the cascade approximation in the medium takes the simple form 
\begin{align}
  \frac{P_{3, \,{\rm med}}^{\rm \, cascade}}{s_{123}} &= 
  \frac{P^{2,1+3}_{gq}P^{1,3}_{gq,{\rm med}} +  P^{2,1+3}_{gq, {\rm med}}P^{1,3}_{gq} }{s_{13}}+\frac{P_{gq}^{1,2+3}P_{gq,{\rm med}}^{2,3}+P_{gq,{\rm med}}^{1,2+3}P_{gq}^{2,3}   }{s_{23}}  +\frac{P_{gq,{\rm med}}^{1+2,3}P_{gg}^{1,2}+P_{gq}^{1+2,3}P_{gg,{\rm med}}^{1,2} }{s_{12}} \, ,
\end{align}
where the subscript med denotes the medium modified splitting functions, which can be directly obtained from the particle distributions computed above.

Using the results in Eq.~\eqref{eq:dN_full_iso}, we can directly study the OPE limit of the in-medium contribution. Noticing that 
\begin{align}
 \frac{d\Sigma^{P(N)}_{\rm no \, vac}}{d R_L} = \int_0^1 dz \, (1-z^N-(1-z)^N) \frac{\alpha_s}{\pi R_L}  \left( \frac{2\pi}{\alpha_s} \P^2\frac{dN_{i\rightarrow jk}}{dzd\P^2}\right)\, ,
\end{align}
where $\P^2 = (\tilde p_t)^2 R_L^2/2$. Power expanding in the angular scale, we find that the OPE coefficients read
\begin{align}
b^{{\rm P}(N)}_{\tau,0} = \int_0^1 dz \, (1-z^N - (1-z)^N) \frac{\alpha_s}{\pi} C_{ji} P_{ji}(z)    \mathcal{W}_\tau \, ,
\end{align}
with 
\begin{align}\label{eq:fun}
 \mathcal{W}_{\tau=2k} = \frac{2}{k!} \left(\frac{s_2}{4 s_1^{3/2}}z(1-z)\right)^k{\rm Re} \left( \left(\frac{-i}{\tilde h_2(\sqrt{s_1})}\right)^k - 4k \int_{X_v^+}\left(\frac{i}{\tilde h_3}\right)^k\tilde h_1^2\left(1 + k\frac{\tilde h_2}{\tilde h_3}\right)\right)\, ,
\end{align}
for $k=2,3,\cdots$, and where the dimensionless coefficients $\tilde h_i$ are given by
\begin{align}
    \tilde h_i = \sqrt{\frac{p_t \sqrt{2}}{\hat q}}h_i|_{(x_v^+,\bar x_v^+) \rightarrow \sqrt{\frac{p_t \sqrt{2}}{\hat q}}(x_v^+,\bar x_v^+)}\,,\qquad \tilde h_2(\sqrt{s_1}) = \tilde h_2|_{\Delta t \rightarrow \sqrt{s_1}} \,, 
\end{align}
and $X_v^+ = \{0 < x_v^+ < \sqrt{s_1}, x_v^+ < \bar x_v^+ < \sqrt{s_1}\}$. The relevant dimensionless scales are $s_1 = \frac{\hat q (L^+)^2}{ \sqrt{2} p_t} = \frac{\omega_c}{ p_t}$ and $s_2 = \hat q (L^+)^3 = 1/\theta_c^2$, as identified in the main text.

\begin{figure}
    \centering
    \includegraphics[width=0.4\linewidth]{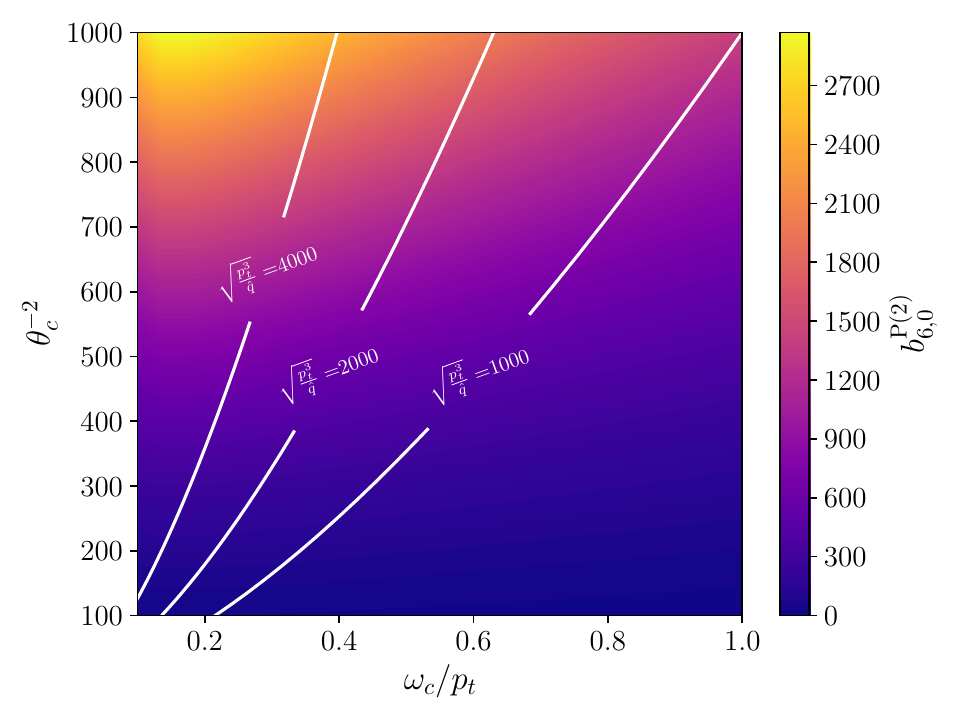}
    \includegraphics[width=0.4\linewidth]{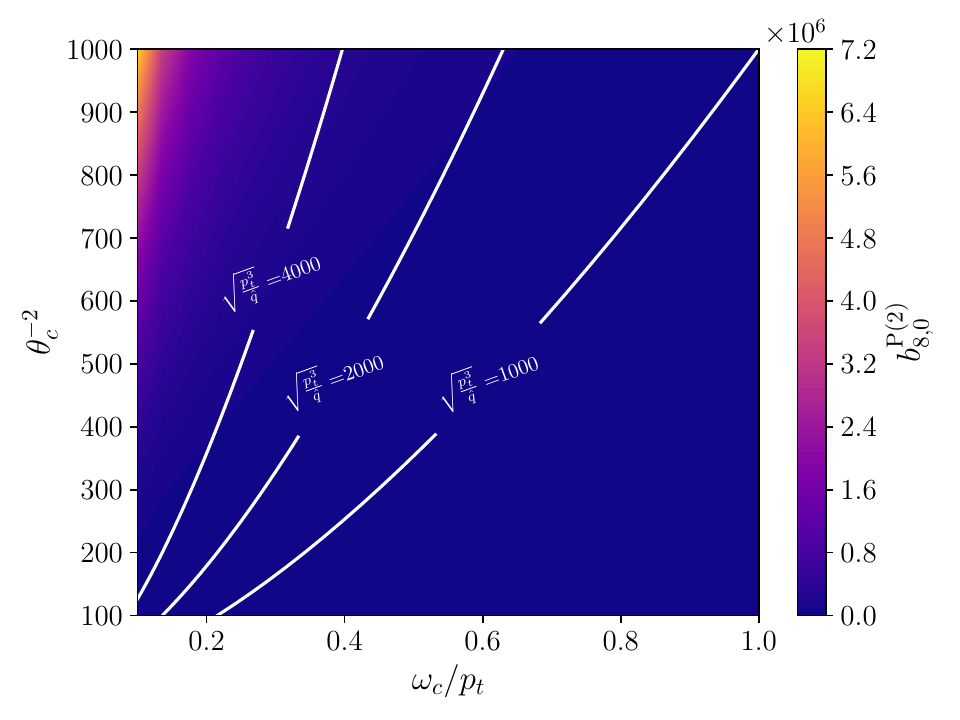}
    \caption{Leading spin, $j=0$, and twist, $\tau=6,8$, OPE coefficients for the E2C distribution following Eq.~\eqref{eq:fun}. }
    \label{fig:wks}
\end{figure}

\subsection{Medium response contribution to ENCs and PENCs}

Here, we briefly review the details of the medium response contribution to the energy-energy correlators, following the construction introduced in~\cite{Barata:2024ieg}. Here, we do not specify the medium model, while dealing with the classical part of the energy flux.

Starting with the E2C, we first fix the fluctuating contribution of harder particles, approximating it, for a moment, with the vacuum-like scaling $\frac{d\Sigma^{(2)}_{\rm P, vac}}{dR_L}\simeq a^{P(2)}_{2,0} R_L^{\gamma(3)-1}$.  Then, in the collinear limit of small $R_L$, we readily find that
\begin{align}
&\frac{d\Sigma^{(2)}}{dR_L}\Bigg\vert_{\rm vac + response}=\frac{1}{p_t^2} \int_{n_{1,2}}\langle \mathcal{E}_{1}\mathcal{E}_{2}\rangle\delta(\cos{R_L}-n_1\cdot n_2)\sin R_L \simeq \frac{a^{P(2)}_{2,0}}{R_L^{1-\gamma(3)}}\notag\\
&\hspace{1.5cm}+\frac{(2\pi)}{p_t^2}\hat{\sum}\left[ \int d\phi_1 d\cos\theta_1\,\mathcal{E}_c(\cos\theta_1)\mathcal{E}_c\left(\cos\theta_1\cos R_L+\sin\theta_1\sin R_L\cos\phi_1\right)\right]\sin R_L\,,
\end{align}
where $\hat{\sum}$ corresponds to an averaging over multiple events (and we keep it implicit in what follows), while the geometric correlation in the given event is made explicit. Further expanding the classical energy flux in $R_L$, we find that all the even terms are zero after azimuthal averaging, as long as the energy flux is symmetric around the jet axis,
\begin{align}
&\frac{d\Sigma^{(2)}}{dR_L}\Bigg\vert_{\rm vac + response}=
\frac{a^{{\rm P}(2)}_{2,0}}{R_L^{1-\gamma(3)}} +\sum_{k=2}c^{{\rm P}(2)}_\tau\left[\mathcal{E}_c\right]R_L^{\tau-3}\Big|_{\tau=2k}
\end{align}
where for $\mathcal{E}_c(\bn)=\frac{\Delta}{\pi\theta_0^2} e^{-\theta^2/\theta_0^2}$ we have $c^{{\rm P}(2)}_4=\Delta^2/p_t^2\theta_0^2$. However, the even powers are non-zero for azimuthally perturbed fluxes, as one can easily check substituting $\mathcal{E}_c=\delta(\phi)f(\theta)$.

Turning to the PE3C, defined with
\begin{align}
  \frac{d\Sigma^{(3)}_{\rm P}}{dR_L} &=\frac{1}{p_t^3} \int_{n_{1,2,3}}\langle \mathcal{E}_1\mathcal{E}_2\mathcal{E}_3\rangle\delta(\cos{R_L}-n_1\cdot n_2)\Theta(|n_1-n_2|-|n_2-n_3|)\Theta(|n_2-n_3|-|n_1-n_3|)\sin R_L\,,
\end{align}
we find that
\begin{align}
\frac{d\Sigma^{(3)}_{\rm P,ccc}}{dR_L} &\simeq \frac{(2\pi)^2}{p_t^3}\,\left[\int d\cos\theta_1\,\mathcal{E}_{s}^3(\theta_1)\right]\,\left(\frac{\pi}{3}-\frac{1}{4}\tan\frac{\pi}{3}\right) R_L^3 = \frac{\Delta^3}{\theta_0^2p_t^3}\,\left(\frac{2}{9}-\frac{\sqrt{3}}{6\pi}\right)R_L^3\,,
\end{align}
where in the last step we have used the explicit form of the model classical energy flux, and
\begin{align}
\frac{d\Sigma^{(3)}_{\rm P,cch}}{dR_L}+\frac{d\Sigma^{(3)}_{\rm P,chc}}{dR_L}+\frac{d\Sigma^{(3)}_{\rm P,hcc}}{dR_L}
&\simeq \frac{ a^{P(3)}_{2,0} \Delta}{6\pi p_t}\,\mathcal{C}\,R_L^{1+\gamma(3)}\,,
\end{align}
where
\begin{align}
\mathcal{C}=2\int_0^{\pi/3}d\phi\int_{\frac{1}{2\cos\phi}}^1dt\left(t+t^{\gamma(3)-1}+\frac{t}{(1+t^2-2t\cos\phi)^{1-\frac{\gamma(3)}{2}}}\right)\,.
\end{align}
Combining these terms, we finally obtain the numerator of Eq.~\eqref{eq:ratio:E3C_E2C}, reading
\begin{align}
\frac{d\Sigma^{(3)}_{\rm P}}{dR_L}\Bigg\vert_{\rm vac + response}
&\simeq \frac{ a^{P(3)}_{2,0}}{R_L^{1-\gamma(4)}} 
+ \frac{ a^{P(3)}_{2,0} \Delta}{6\pi p_t}\,\mathcal{C}\,R_L^{1+\gamma(3)} + \frac{\Delta^3}{\theta_0^2p_t^3}\,\left(\frac{2}{9}-\frac{\sqrt{3}}{6\pi}\right)R_L^3\,.
\end{align}

\end{document}